\documentclass{article}

\newcommand{\p}[1]{(\ref{#1})}

\newcommand{\bQ}{{\overline Q}{}}
\newcommand{\bS}{{\overline S}{}}

\newcommand{\bpsi}{{\bar\psi}{}}

\newcommand{\tJ}{{\widetilde J}}
\newcommand{\whJ}{{\widehat J}}

\newcommand{\be}{\begin{equation}}
\newcommand{\ee}{\end{equation}}
\newcommand{\bea}{\begin{eqnarray}}
\newcommand{\eea}{\end{eqnarray}}

\newcommand{\ba}{\begin{array}} \newcommand{\ea}{\end{array}}

\def\im{{\rm i}}

\newcommand{\nn}{\nonumber}

\usepackage{amscd,amsmath,amssymb}

\begin{document}
\thispagestyle{empty}
\vspace{2cm}
\begin{flushright}
\end{flushright}\vspace{2cm}
\begin{center}
{\Large\bf A note on $SU(1,1|n)$ and $OSp(6|2)$ superconformal mechanics}
\end{center}
\vspace{1cm}
\begin{center}
{\large\bf  Nikolay Kozyrev${}^a$}
\end{center}

\vspace{0.2cm}

\begin{center}
{ \it ${}^a$
Bogoliubov  Laboratory of Theoretical Physics, JINR,
141980 Dubna, Russia}
\end{center}
\vspace{0.5cm}
\begin{center}
{\tt nkozyrev@theor.jinr.ru}
\end{center}
\vspace{2cm}

\begin{abstract}\noindent
In this article we consider the construction of the superconformal mechanics that realize $SU(1,1|n)$ and $OSp(6|2)$ symmetries and involve interactions with non-Abelian bosonic currents. If is shown that for $N>4$ supersymmetries the currents involved have to satisfy the algebraic equations. General considerations on methods of solving these equations are given. In the obtained particular solutions currents are expressed in terms of semi-dynamical variables (harmonics), and, on one instance, coordinates and momenta.
\end{abstract}

\setcounter{page}{1}
\setcounter{equation}{0}

\section{Introduction}
The mechanical systems possessing superconformal symmetries have been under study for many years. Recent interest in this subject is induced by the newly obtained solutions of $10$ and $11$-dimensional supergravities that preserve some supersymmetry and are products of $AdS_2$ and a compact manifold, such as a sphere or complex projective space (some examples can be found in \cite{N3_1,N3_2,N4_1,N4_2}). This suggests the existence of the superconformal mechanics which Hamiltonian is a Hamiltonian on the appropriate compact manifold, modified by the interaction with the dilaton.

Most of the studies of superconformal mechanics are dedicated to the mechanics with $N=4$, and, to lesser extent, $N=8$ supersymmetry, although superconformal algebras in one dimension with any number of supercharges are known \cite{Dict}. Indeed, the superconformal mechanics with $N=3$ \cite{kk1} and $N=7$ supersymmetries \cite{toppan1,kn3} were constructed. Moreover, the system with $su(1,1|n)$ superconformal symmetry with nontrivial potential and interaction with fermions was known already in 1989 \cite{IKL}.

Among the interactions that superconformal mechanics can possess is, besides the dilaton potential, the coupling to the non-Abelian current that usually appears in the (anti)commutation relations of the superconformal algebra. Such systems with $N=4$ and $N=8$ supersymmetry and with bosonic and fermionic currents were constructed using both Hamiltonian and superfield formalism. In the latter case, bosonic currents are typically constructed in terms semi-dynamic variables called harmonics,\footnote{For example, see \cite{FI5,FI4}.} although it is not necessary and currents can also be constructed in terms of dynamical variables and their momenta, as was done in \cite{kk1} and, implicitly, in \cite{fi1}, where harmonics can be covariantly split into real quadruplets that can be treated as coordinates and momenta to each other \cite{KKN1}.

Analyzing the known supersymmetric mechanics with bosonic non-Abelian currents within the Hamiltonian formalism, one can note that for systems with $N=4$ supersymmetry the supercharges satisfy superconformal algebra relations without reference to the structure of currents. In contrast, the algebra of $OSp(8|2)$ mechanics \cite{kn1} closes only for a given parametrization of currents. Therefore, it would be interesting to study the superconformal systems with number of supersymmetries different from $4$ or $8$. Such studies were already performed in \cite{GL1}, where it was shown that currents in $su(1,1|n)$ system have to satisfy an unsolved algebraic equation, and in \cite{Cher1,Cher2}. In the latter works nonlinear realizations of $SU(1,1|n)$ and $OSp(n|2)$ were considered and the physical fermionic fields were introduced as parameters accompanying the supercharges, suggesting that supersymmetry in these systems is spontaneously broken.

In this article, we reexamine the $SU(1,1|n)$ superconformal systems and find solutions of the related algebraic equation. Additionally, we construct a system that possesses $OSp(6|2)$ superconformal symmetry and provide considerations on the solution of equations that appear in such models.

\section{$SU(1,1|n)$ superconformal mechanics}
The systems, possessing $SU(1,1|n)$ symmetry, can be constructed using the Hamiltonian formalism. Let us consider dilaton $r$, its momentum $p_r$, $u(n)$ generators $J_\alpha{}^\beta = \big( J_\beta{}^\alpha  \big){}^\dagger$ and fermionic fields $\psi_\alpha$, $\bpsi{}^\alpha = \big(  \psi_\alpha \big)^\dagger$. Their Dirac brackets read
\be\label{prrJpsi}
\big\{ p_r, r \big\} =1, \;\; \big\{ J_\alpha{}^\beta, J_\mu{}^\nu  \big\} =\im \delta_\mu{}^\beta J_\alpha{}^\nu- \im \delta_\alpha{}^\nu J_\mu{}^\beta,  \;\; \big\{ \psi_\alpha, \bpsi{}^\beta \big\} = \im \delta_\alpha{}^\beta.
\ee
Now one can consider supercharges and Hamiltonian with structure typical to superconformal mechanics
\bea\label{suQH}
Q_\alpha = p_r \psi_\alpha +\frac{\im}{r} \big( a_1 J_\alpha{}^\mu \psi_\mu  +a_2 J_\mu{}^\mu \psi_\alpha +  a_3 \psi_\alpha\, \psi_\mu \bpsi{}^\mu    \big), \;\; \nn \\
\bQ{}^\alpha = p_r \bpsi{}^\alpha -\frac{\im}{r} \big( {\bar a}_1 J_\mu{}^\alpha \bpsi{}^\mu+{\bar a}_2 J_\mu{}^\mu \bpsi{}^\alpha + {\bar a}_3 \bpsi{}^\alpha\, \psi_\mu \bpsi{}^\mu    \big), \nn \\
H = \frac{1}{2}p_r^2 + b_1 \frac{p_r}{r}J_\mu{}^\mu + \frac{1}{r^2}\left( b_2 J_\mu{}^\nu J_\nu{}^\mu +b_3 \big( J_\mu{}^\mu \big)^2 + b_4 J_\mu{}^\nu \,\psi_\nu \,\bpsi{}^\mu + b_5 J_\mu{}^\mu \,\psi_\nu \,\bpsi{}^\nu +b_6 \big( \psi_\nu \,\bpsi{}^\nu  \big)^2 \right).
\eea
The (super)conformal generators are taken as
\be\label{suDKS}
D= \frac{1}{2}r p_r, \;\; K= \frac{1}{2}r^2, \;\; S_\alpha = r\psi_\alpha, \;\; \bS{}^\alpha = r \bpsi{}^\alpha.
\ee
As can be checked, the fermionic objects $\whJ_\alpha{}^\beta = \psi_\alpha \, \bpsi{}^\beta$ form $u(n)$ algebra similar to \p{prrJpsi}. Therefore, we consider $\tJ_\alpha{}^\beta = J_\alpha{}^\beta + \whJ_\alpha{}^\beta$ to be $R$-symmetry generators.

Many of $su(1,1|n)$ superalgebra relations are satisfied due to choice of ansatz \p{suQH} without imposing any constraints:
\bea\label{SUrels1}
\big\{ D, H \big\} =-H, \;\; \big\{ D, K \big\} =K, \;\; \big\{ K,Q_\alpha  \big\} = -S_\alpha, \;\;  \big\{ K,\bQ{}^\alpha  \big\} = -\bS{}^\alpha,\;\; \big\{ S_\alpha, \bS{}^\beta \big\} = 2\im \delta_\alpha{}^\beta K, \;\;  \nn \\
 \big\{ D,Q_\alpha  \big\} = -\frac{1}{2}Q_\alpha, \;\;  \big\{ D,\bQ{}^\alpha  \big\} = -\frac{1}{2}\bQ{}^\alpha, \;\; \big\{ D,S_\alpha  \big\} = \frac{1}{2}S_\alpha, \;\;  \big\{ D,\bS{}^\alpha  \big\} = \frac{1}{2}\bS{}^\alpha, \nn \\
\big\{ \tJ_\alpha{}^\beta , \tJ_{\mu}{}^\nu   \big\} =\im \delta_\mu{}^\beta  \tJ_\alpha{}^\nu-\im \delta_\alpha{}^\nu  \tJ_\mu{}^\beta, \;\; \big\{ \tJ_\alpha{}^\beta, Q_{\mu}  \big\} = \im \delta_\mu{}^\beta Q_{\alpha}, \;\; \big\{ \tJ_\alpha{}^\beta, S_{\mu}  \big\} = \im \delta_\mu{}^\beta S_{\alpha}, \nn \\ \big\{ \tJ_\alpha{}^\beta , \bQ{}^{\mu}  \big\} = -\im \delta_\alpha{}^\mu \bQ{}^{\beta}, \;\; \big\{ \tJ_\alpha{}^\beta, \bS{}^{\mu}  \big\} =-\im \delta_\alpha{}^\mu \bS{}^{\beta}.
\eea
The relations
\bea\label{SUrels2}
\big\{ Q_\alpha, \bS{}^\beta \big\} = 2\im \delta_\alpha{}^\beta D -\delta_\alpha{}^\beta \tJ_\mu{}^\mu +2 \tJ_\alpha{}^\beta , \;\;
\big\{ \bQ{}^\beta, S_\alpha \big\} = 2\im \delta_\alpha{}^\beta D + \delta_\alpha{}^\beta \tJ_\mu{}^\mu-2 \tJ_\alpha{}^\beta
\eea
demand that $a_1 = {\bar a}_1 =-2$, $a_3 ={\bar a}_3 =1$, and, if $J_\mu{}^\mu \neq 0$, $a_2 = {\bar a}_2 =1$, effectively fixing the supercharges and implying further relations
\be\label{SUrels3}
\big\{ Q_\alpha, Q_\beta  \big\} =0, \;\; \big\{ \bQ{}^\alpha, \bQ{}^\beta  \big\}=0,\;\; \big\{ Q_\alpha, S_\beta  \big\}=0, \;\; \big\{ \bQ{}^\alpha , \bS{}^\beta  \big\}=0.
\ee
The brackets
\be\label{SUrels4}
\big\{  H, K \big\} =2D, \;\; \big\{ H,S_\alpha  \big\} = Q_\alpha, \;\;  \big\{ H,\bS{}^\alpha  \big\} = \bQ{}^\alpha
\ee
imply that  $b_4=-2$, $b_5=1$, $b_6= \frac{1}{2}$ and $b_1=0$ (if $J_\mu{}^\mu \neq 0$), with $b_2$, $b_3$ remaining free.

With supercharges completely fixed, one can calculate the final nontrivial algebra bracket $\big\{ Q_\alpha, \bQ{}^\beta \big\}$ to find that it contains explicitly non-diagonal terms:
\be\label{suQQb}
\big\{ Q_\alpha, \bQ{}^\beta \big\} = \frac{4\im}{r^2} J_\alpha{}^\mu J_\mu{}^\beta - \frac{4\im }{r^2} J_\alpha{}^\beta J_\mu{}^\mu+ \delta_\alpha^\beta \ldots
\ee
At this time, it is not enough to choose constants properly and algebra relations can be satisfied if an equation is imposed
\be\label{suJeq}
J_\alpha{}^\mu J_\mu{}^\beta - J_\alpha{}^\beta J_\mu{}^\mu = \delta_\alpha{}^\beta \big( c_1 J_\mu{}^\nu J_\nu{}^\mu + c_2 \big(J_\mu{}^\mu \big)^2 \big).
\ee
In the particular case of $n=2$ (four real supercharges), left hand side of \p{suJeq}  is proportional to $\delta_{\alpha}{}^\beta$ identically and one can freely choose representation $J_\alpha{}^\beta$ belongs to. For $n\geq 3$ (six or more real supercharges), \p{suJeq} is an algebraic equation that strongly constrains $J_\alpha{}^\beta$.  Postponing solving this equation to the Section 4, one can note that simplest possible solution is
\be\label{suJsol}
J_\alpha{}^\beta = v_\alpha {\bar v}{}^\beta, \;\; c_1 =c_2 =b_2=0, \;\; b_3=\frac{1}{2}.
\ee
As $J_\alpha{}^\beta$ are to form $u(n)$ algebra, $v_\alpha$, ${\bar v}{}^\beta$ should have brackets
\be\label{suvvb}
\big\{  v_\alpha, {\bar v}{}^\beta  \big\} = -\im \delta_\alpha{}^\beta.
\ee
If equation \p{suJeq} is satisfied, $\big\{Q_\alpha, \bQ{}^\beta  \big\}=2\im \delta_\alpha^\beta H$ with
\bea\label{su11N}
Q_\alpha = p_r \psi_\alpha +\frac{\im}{r} \big( -2 J_\alpha{}^\mu \psi_\mu  +J_\mu{}^\mu \psi_\alpha +  \psi_\alpha\, \psi_\mu \bpsi{}^\mu    \big), \;\; \nn \\
\bQ{}^\alpha = p_r \bpsi{}^\alpha +\frac{\im}{r} \big( 2 J_\mu{}^\alpha \bpsi{}^\mu  - J_\mu{}^\mu \bpsi{}^\alpha -  \bpsi{}^\alpha\, \psi_\mu \bpsi{}^\mu    \big), \nn \\
H = \frac{1}{2}p_r^2 + \frac{2c_1}{r^2}J_\mu{}^\nu J_\nu{}^\mu+\frac{1+4c_2}{2r^2}\big(J_\mu{}^\mu\big)^2 - \frac{2}{r^2}J_\mu{}^\nu \psi_\nu\,\bpsi{}^\mu + \frac{1}{r^2}J_\mu^\mu \, \psi_\nu \bpsi{}^\nu +\frac{1}{2r^2}\big(\psi_\nu \bpsi{}^\nu   \big)^2.
\eea
It is worth noting that Hamiltonian \p{su11N} can be written in terms of Casimir operators of bosonic, fermionic and complete $u(n)$ algebras. In general, two quadratic Casimir operators exist for each algebra, $J_\mu{}^\nu J_\nu{}^\mu$, $\big( J_\mu{}^\mu \big)^2$ and analogs, though for the fermionic algebra they are similar up to a sign. In terms of these five independent operators, the Hamiltonian reads
\bea\label{su11NH}
H = \frac{1}{2}p_r^2 + \frac{1}{r^2} \left( (1+2 c_1)C_{bos} +2 c_2 {\cal C}_{bos} - C + \frac{1}{2}{\cal C} + C_{ferm}   \right), \nn \\
C_{bos} = J_\mu{}^\nu J_\nu{}^\mu, \;\; {\cal C}_{bos} = \big( J_\mu{}^\mu \big)^2, \;\; C=  \tJ_\mu{}^\nu \tJ_\nu{}^\mu, \;\; {\cal C} = \big( \tJ_\mu{}^\mu \big)^2, \;\;  C_{ferm}=\whJ_\mu{}^\nu \whJ_\nu{}^\mu.
\eea

\section{$OSp(6|2)$ superconformal mechanics}
The $OSp(6|2)$ mechanics can be constructed in a similar way to the $SU(1,1|n)$ case. Let us note that for $osp(n|2)$ the internal symmetry algebra is $so(n)$, and supercharges and conformal supercharges transform as vectors with respect to it. The cases of $osp(4|2)$ and $osp(6|2)$ are special, as the internal symmetry algebras $so(4)\sim su(2)\times su(2)$ and  $so(6)\sim su(4)$ can be treated as unitary. The $OSp(4|2)$ case is already thoroughly studied \cite{FI5,kk1}, and we, therefore, concentrate on $OSp(6|2)$. Isomorphism $so(6)\sim su(4)$ suggests writing the algebra and supercharges in $SU(4)$ spinor notation, where $SO(6)$ vectors are antisymmetric spin-tensors $\psi_{[\alpha\beta]}$, $\alpha=1,2,3,4$; these antisymmetric pairs can be raised and lowered with totally antisymmetric $\epsilon_{\alpha\beta\mu\nu}$ symbol, which also allows to define reality condition for $\psi^{\alpha\beta}$
\be\label{osp6eps}
\psi{}^{[\alpha\beta]} = \frac{1}{2}\epsilon^{\alpha\beta\mu\nu}\psi_{\mu\nu}, \;\; \psi{}_{[\alpha\beta]} = \frac{1}{2}\epsilon_{\alpha\beta\mu\nu}\psi{}^{\mu\nu}, \;\; \big(\psi^{\alpha\beta}\big)^\dagger = \frac{1}{2}\epsilon_{\alpha\beta\mu\nu}\psi^{\mu\nu}.
\ee
The $su(4)$ generators in this formalism are usual $J_\alpha{}^\beta$, with $J_{\alpha}{}^{\alpha}=0$.

It is easy to write down the ansatz for the supercharges, involving dilaton $r$, its momentum $p_r$, $su(4)$ generators and fermions $\psi{}^{[\alpha\beta]}$ with brackets
\be\label{osp6xpJpsi}
\big\{ p_r,r  \big\} =1, \;\; \big\{  J_\alpha{}^{\beta}, J_{\mu}{}^{\nu} \big\} = \im \delta_{\mu}{}^{\beta}J_{\alpha}{}^{\nu} - \im \delta_\alpha{}^\nu J_{\mu}{}^{\beta}, \;\; \big\{  \psi^{\alpha\beta}, \psi^{\mu\nu} \big\} = \im \epsilon^{\alpha\beta\mu\nu}.
\ee
The ansatz for $Q^{\alpha\beta}$ contains no terms cubic in the fermions, which is typical to $osp(n|2)$ mechanics. Remarkably, it involves only one free parameter:
\be\label{osp6Q}
Q^{\alpha\beta} = p_r \psi^{\alpha\beta} + \frac{\im a}{2r}\big( J_\rho^{\alpha}\psi^{\beta\rho} - J_{\rho}{}^{\beta}\psi^{\alpha\rho}   \big), \;\; \big( Q^{\alpha\beta} \big)^\dagger = Q_{\alpha\beta} = \frac{1}{2}\epsilon_{\alpha\beta\mu\nu}Q^{\mu\nu}.
\ee
Evaluating the bracket of the supercharges, one finds
\bea\label{osp6QQ}
\big\{ Q^{\alpha\beta}, Q^{\mu\nu}\big\}  = \epsilon^{\alpha\beta\mu\nu} \left(  \im  p_r^2  - \im \frac{a^2}{4r^2} J_{\rho}{}^{\sigma}\psi_{\sigma\lambda}\psi{}^{\rho\lambda}\right) +\frac{\im}{r^2}\psi^{\alpha\beta} J_{\rho}{}^{[\mu} \psi{}^{\nu]\rho} \left( -a -\frac{a^2}{2}   \right) +\frac{\im}{r^2}\psi^{\mu\nu} J_{\rho}{}^{[\alpha} \psi{}^{\beta]\rho} \left( -a -\frac{a^2}{2}   \right)-\nn \\
-\frac{\im a^2}{4r^2} \left( -J_{\rho}{}^{\alpha}J_{\sigma}{}^{\mu}\epsilon^{\beta\nu\rho\sigma} + J_{\rho}{}^{\beta}J_{\sigma}{}^{\mu}\epsilon^{\alpha\nu\rho\sigma}  +J_{\rho}{}^{\alpha}J_{\sigma}{}^{\nu}\epsilon^{\beta\mu\rho\sigma} - J_{\rho}{}^{\beta}J_{\sigma}{}^{\nu}\epsilon^{\alpha\mu\rho\sigma}     \right).
\eea
The undesirable terms, containing $\psi^2$, can be cancelled by choosing $a=-2$. The remaining bosonic terms should be proportional to $\epsilon^{\alpha\beta\mu\nu}$. As one can guess, the $su(4)$ generator constructed from harmonics
\be\label{osp6Jsol}
J_{\alpha}{}^\beta =  v_\alpha {\bar v}{}^\beta - \frac{1}{4}\delta_\alpha{}^\beta v_\mu {\bar v}{}^\mu, \;\; \big\{ v_\alpha, {\bar v}{}^{\beta}\big\} = -\im \delta_{\alpha}{}^{\beta}
\ee
satisfies the necessary condition:
\be\label{osp6condJ}
 -J_{\rho}{}^{\alpha}J_{\sigma}{}^{\mu}\epsilon^{\beta\nu\rho\sigma} + J_{\rho}{}^{\beta}J_{\sigma}{}^{\mu}\epsilon^{\alpha\nu\rho\sigma}  +J_{\rho}{}^{\alpha}J_{\sigma}{}^{\nu}\epsilon^{\beta\mu\rho\sigma} - J_{\rho}{}^{\beta}J_{\sigma}{}^{\nu}\epsilon^{\alpha\mu\rho\sigma} = - \frac{1}{4}\big( v_\rho {\bar v}^{\rho}   \big)^2\epsilon^{\alpha\beta\mu\nu}.
\ee
Therefore, in this case
\be\label{osp6QQH}
\big\{ Q^{\alpha\beta}, Q^{\mu\nu}\big\}  = 2\im \epsilon^{\alpha\beta\mu\nu}H, \;\; H = \frac{1}{2}p_r^2 + \frac{1}{8r^2} \big( v_\mu {\bar v}^{\mu}   \big)^2 - \frac{1}{2r^2}v_\alpha {\bar v}{}^\beta \psi_{\lambda\beta}\psi{}^{\lambda\alpha}.
\ee
Together with generators
\be\label{osp6DKShJ}
D= \frac{1}{2}r p_r, \;\; K =\frac{1}{2}r^2, \;\; S^{\mu\nu} = r \psi^{\mu\nu}, \;\; \tJ_\alpha{}^\beta = J_\alpha{}^\beta+ \frac{1}{2}\psi_{\alpha\lambda}\psi{}^{\beta\lambda}
\ee
supercharges $Q^{\alpha\beta}$ form $osp(6|2)$ superalgebra
\bea\label{osp6alg}
\big\{ Q^{\alpha\beta}, Q^{\mu\nu}\big\}  = 2\im \epsilon^{\alpha\beta\mu\nu}H, \;\; \big\{ S^{\alpha\beta}, S^{\mu\nu}\big\}  = 2\im \epsilon^{\alpha\beta\mu\nu}K, \nn \\
\big\{ Q^{\alpha\beta}, S^{\mu\nu}\big\}  = 2\im \epsilon^{\alpha\beta\mu\nu}D -\frac{1}{2}\left( \epsilon^{\alpha\mu\nu\rho}\tJ_\rho{}^\beta - \epsilon^{\beta\mu\nu\rho}\tJ_\rho{}^\alpha  + \epsilon^{\alpha\beta\nu\rho}\tJ_\rho{}^\mu  -\epsilon^{\alpha\beta\mu\rho}\tJ_\rho{}^\nu       \right), \nn \\
\big\{\tJ_\alpha{}^\beta , Q^{\mu\nu}\big\} = \frac{\im}{2}\delta_\alpha{}^\beta Q{}^{\mu\nu} -\im \delta_\alpha{}^\mu Q^{\beta\nu} - \im \delta_{\alpha}{}^\nu Q^{\mu\beta}, \;\;
\big\{ \tJ_\alpha{}^\beta , S^{\mu\nu}\big\} = \frac{\im}{2}\delta_\alpha{}^\beta S{}^{\mu\nu} -\im \delta_\alpha{}^\mu S^{\beta\nu} -\im \delta_{\alpha}{}^\nu S^{\mu\beta}, \nn \\
\big\{ \tJ_\alpha{}^\beta, \tJ_\mu{}^\nu \big\}  = \delta_{\mu}{}^{\beta}  \tJ_\alpha{}^\nu -\delta_{\alpha}{}^\nu  \tJ_\mu{}^\beta, \;\;
\big\{ D,K \big\} =K, \;\; \big\{ D,H\} = -H, \;\; \big\{  H,K \big\} = 2D, \nn \\
\big\{ D,Q^{\alpha\beta}\big\} =-\frac{1}{2}Q^{\alpha\beta}, \;\; \big\{ D,S^{\alpha\beta}\big\} =\frac{1}{2}S^{\alpha\beta}, \;\; \big\{ K,Q^{\alpha\beta}\big\} =-S^{\alpha\beta}, \;\; \big\{ H,S^{\alpha\beta}\big\} =Q^{\alpha\beta}.
\eea
In terms of the Casimir operators of the bosonic and complete $su(4)$ algebras
\be\label{so6cas}
C_{su(4),bos} =J_\mu{}^\nu J_\nu{}^\mu = \frac{3}{4}\big( v_\mu \bar{v}^\mu \big)^2, \;\; C_{su(4)} =\tJ_\mu{}^\nu \tJ_\nu{}^\mu = \frac{3}{4}\big( v_\mu \bar{v}^\mu \big)^2 + v_\alpha \bar{v}^\beta \psi_{\beta\lambda}\psi^{\alpha\lambda}
\ee
the Hamiltonian \p{osp6QQH} can be written as
\be\label{osp6Ham}
H = \frac{1}{2}p_r^2 + \frac{1}{r^2} \left( \frac{2}{3} C_{su(4),bos} -\frac{1}{2} C_{su(4)} \right).
\ee

Let us note that one can use $SO(6)$ vector notation instead of $SU(4)$ spinor one. In this case, $\psi^{[\alpha\beta]}$ would correspond to the $SO(6)$ vector $\psi{}_a$, $a=1,\ldots,6$,\footnote{$\gamma$-matrices, relating them, can be taken as usual Lorentzian $\gamma$-matrices in six dimensions, with $\gamma_6 = \im \gamma_0$.} and traceless bispinor $J_\alpha{}^\beta$ to the antisymmetric tensor $J_{ab}$. The ansatz for the supercharges and the constraint on the generators $J_{ab}$ turn out to be very simple,
\be\label{osp6so6}
Q_a = p_r \psi_a + \frac{\alpha}{r}J_{ab}\psi_b, \;\; \big\{Q_a, Q_b\big\}=2\im \delta_{ab}H \;\; \mbox{if} \;\; J_{ac}J_{bc} \sim \delta_{ab}.
\ee
(Similar relations appear in the $osp(n|2)$ case). The constraint \p{osp6so6}, however, it is not easy to solve. Solutions of analogous relations are known in the $osp(4|2)$ and $osp(8|2)$ cases \cite{kn3}, but solution with six supersymmetries turns out to be substantially different. Let us show how it can be found in the next Section.

\section{Solving algebraic equation on currents}
Both $su(1,1|n)$ and $osp(6|2)$ cases involve current $J_\alpha{}^\beta$, which has to satisfy algebraic equations \p{suJeq}, \p{osp6condJ},
\be\label{suJeq2}
J_\alpha{}^\mu J_\mu{}^\beta - J_\alpha{}^\beta J_\mu{}^\mu = \delta_\alpha{}^\beta A
\ee
in the $su(1,1|n)$ case and
\be\label{osp6condJ2}
J_\mu{}^\mu=0, \;\;  -J_{\rho}{}^{\alpha}J_{\sigma}{}^{\mu}\epsilon^{\beta\nu\rho\sigma} + J_{\rho}{}^{\beta}J_{\sigma}{}^{\mu}\epsilon^{\alpha\nu\rho\sigma}  +J_{\rho}{}^{\alpha}J_{\sigma}{}^{\nu}\epsilon^{\beta\mu\rho\sigma} - J_{\rho}{}^{\beta}J_{\sigma}{}^{\nu}\epsilon^{\alpha\mu\rho\sigma} = B \epsilon^{\alpha\beta\mu\nu}
\ee
in the $osp(6|2)$ case, with particular solutions given by \p{suJsol}, \p{osp6Jsol}. Let us show how the solutions to these equations can be found.

The second equation turns out to be simpler to analyze. As $J_\alpha{}^\beta$ is a Hermitean matrix, it can be diagonalized by the action of unitary matrices:
\be\label{hermmatr}
J_\alpha{}^\beta = U_\alpha{}^\mu D_\mu{}^\nu \big( U^{\dagger}   \big)_\nu{}^\beta, \;\; U_\alpha{}^\mu \big( U^{\dagger}   \big)_\mu{}^\beta = \delta_\alpha^\beta, \;\; D = \mbox{diag}\big(\lambda_1, \lambda_2, \ldots \big).
\ee
In the particular case of $osp(6|2)$ $D_\alpha{}^\beta$ is a traceless $4\times 4$ matrix and should be taken as $D = \mbox{diag}\big( \lambda_1 , \lambda_2, \lambda_3 , - \lambda_1 - \lambda_2-\lambda_3\big)$.
Substituting \p{hermmatr} to \p{osp6condJ2}, one can factor out $U_\alpha{}^\mu$ due to relation
\be\label{Unitrel}
U_{\rho}{}^{\lambda}U_{\sigma}{}^{\tau}\epsilon^{\beta\nu\rho\sigma} = \det U\, \big( U^\dagger  \big)_{\rho}^\beta\, \big( U^\dagger  \big)_{\sigma}^{\mu}\, \epsilon^{\lambda\tau\rho\sigma},
\ee
which follows from unitarity of $U_\alpha{}^\beta$, and find
\bea\label{osp6condJ3}
-D_{\rho}{}^{\alpha}D_{\sigma}{}^{\mu}\epsilon^{\beta\nu\rho\sigma} + D_{\rho}{}^{\beta}D_{\sigma}{}^{\mu}\epsilon^{\alpha\nu\rho\sigma}  +D_{\rho}{}^{\alpha}D_{\sigma}{}^{\nu}\epsilon^{\beta\mu\rho\sigma} - D_{\rho}{}^{\beta}D_{\sigma}{}^{\nu}\epsilon^{\alpha\mu\rho\sigma} = B \epsilon^{\alpha\beta\mu\nu}\;\; \Rightarrow \nn \\
-B = \big(\lambda_1 +\lambda_2\big)^2 = \big(\lambda_1 +\lambda_3\big)^2 = \big(\lambda_2 +\lambda_3\big)^2.
\eea
The latter equations can also be written as
\bea\label{osp6condJ4}
\big( \lambda_1  -\lambda_2   \big)\big( \lambda_1 +\lambda_2 +2 \lambda_3  \big)=0, \nn \\
\big( \lambda_1  -\lambda_3   \big)\big( \lambda_1 +\lambda_3 +2 \lambda_2  \big)=0,  \\
\big( \lambda_2  -\lambda_3   \big)\big( \lambda_2 +\lambda_3 +2 \lambda_1  \big)=0. \nn
\eea
Assuming all right multiples are zero results in $ \lambda_1=\lambda_2 = \lambda_3=0 $, so nontrivial solution has to involve at least two equal $\lambda$'s. Without loss of generality, one can assume that $\lambda_2 =\lambda_3$. Then other equations lead to one
\be\label{osp6condJ5}
\big( \lambda_1  -\lambda_2   \big)\big( \lambda_1 +3 \lambda_2  \big)=0.
\ee
Assuming the first multiple is zero leads to $\lambda_1=\lambda_2 =\lambda_3$. The other possible solution $\lambda_1 = -3\lambda_2$ seems to be different, but actually leads to a matrix with three similar eigenvalues, as in this case $- \lambda_1 - \lambda_2-\lambda_3 =\lambda_2 $, and thus can be discarded. After substitution of the first solution into \p{hermmatr}, one finds that $J_\alpha{}^\beta$ can be expressed in terms of $U_\alpha{}^4 \big( U^{\dagger}   \big)_4{}^\beta$ due to
\bea\label{U4rel}
U_\alpha{}^1 \big( U^{\dagger}   \big)_1{}^\beta+U_\alpha{}^2 \big( U^{\dagger}   \big)_2{}^\beta+U_\alpha{}^3 \big( U^{\dagger}   \big)_3{}^\beta+U_\alpha{}^4 \big( U^{\dagger}   \big)_4{}^\beta=\delta_\alpha^\beta \Rightarrow \nn \\
J_\alpha{}^\beta = -4 \lambda_1  U_\alpha{}^4 \big( U^{\dagger}   \big)_4{}^\beta +\lambda_1 \delta_\alpha^\beta.
\eea
Formally identifying $v_\alpha = \sqrt{-2\lambda_1} U_\alpha{}^4$, $\bar{v}{}^\alpha = \sqrt{-2\lambda_1} \big( U^{\dagger}   \big)_4{}^\alpha$, one obtains solution \p{osp6Jsol}
\be\label{osp6condJ6}
J_\alpha{}^\beta  = v_\alpha \, \bar{v}{}^\beta - \frac{1}{4}\delta_\alpha^\beta v_\gamma \, \bar{v}{}^\gamma.
\ee
The generators \p{osp6condJ6} form $su(4)\sim so(6)$ algebra if one assumes that $\big\{  v_\alpha, \bar{v}{}^\beta \big\}=-\im \delta_\alpha{}^\beta$.

The solution of \p{suJeq2} is more subtle. It is easy to find a solution in the particular case
\be\label{suJeq3}
J_\alpha{}^\mu J_\mu{}^\beta - J_\alpha{}^\beta J_\mu{}^\mu = 0.
\ee
Substituting $J_\alpha{}^\beta$ as \p{hermmatr}, one again factorizes unitary matrices and obtains equation
\be\label{suJeq4}
D_\alpha{}^\mu D_\mu{}^\beta = D_\alpha{}^\beta D_\mu{}^\mu \;\; \Rightarrow \;\; \lambda_i^2 =\lambda_i \sum_k \lambda_k.
\ee
Therefore, either $\lambda_i =0$ or $\lambda_i = \sum_k \lambda_k$, and only one of $\lambda$ can be different from zero; without loss of generality, one can assume that $\lambda_1 \neq 0$. Therefore,
\be\label{suJeq5}
J_\alpha{}^\beta = U_\alpha{}^\mu D_\mu{}^\nu \big( U^{\dagger}   \big)_\nu{}^\beta = \lambda_1 U_\alpha{}^1 \big( U^{\dagger}   \big)_1{}^\beta = v_\alpha{}\bar{v}{}^\beta, \;\; v_\alpha = \sqrt{\lambda_1}U_\alpha{}^1, \;\; \bar{v}{}^\alpha=\sqrt{\lambda_1}\big(U^\dagger \big)_1{}^\alpha =\big( v_\alpha \big)^\dagger.
\ee
Harmonic structure that appears in the solutions \p{osp6condJ6}, \p{suJeq4} is already familiar and is frequently encountered in the superfield approach. Equations \p{suJeq3} and \p{osp6condJ2}, therefore, provide an algebraic explanation behind this structure.

If $A\neq 0$ in \p{suJeq2}, one can substitute $L_{\alpha}{}^{\beta} = J_\alpha{}^\beta - \frac{1}{2}\delta_\alpha{}^\beta J_\gamma{}^\gamma$ into \p{suJeq2} to find that $L_\alpha{}^\gamma L_{\gamma}{}^\beta \sim \delta_\alpha{}^\beta$. Therefore, $L_{\alpha}{}^\beta$ has two eigenvalues with opposite signs. If one of these eigenvalues is encountered only once, $L_{\alpha}{}^{\beta}$, and, consequently, $J_\alpha{}^\beta$ have to be expressed in terms of one set of harmonics, and one obtains a solution extending \p{suJeq5}
\be\label{suJeq6}
J_\alpha{}^\beta = v_\alpha{}\bar{v}{}^\beta + k \delta_\alpha{}^\beta v_\mu \bar{v}{}^\mu.
\ee
If the lowest multiplicity among the eigenvalues is greater than one, similarly greater number of harmonics $v_\alpha{}^i$, $\bar{v}_j^\beta$ appear. They  have to satisfy $v_\alpha{}^i \bar{v}_j^\alpha \sim \delta_j^i$, as they originate from the unitary matrix.

The solution with many harmonics is not satisfactory from different point of view. As $J_\alpha{}^\beta$ have to produce $u(n)$ algebra with respect to the Dirac bracket, one has to define brackets of $v_\alpha{}^i$, $\bar{v}_j^\beta$ that result in the proper algebra of $J_\alpha{}^\beta$, satisfy Jacobi identity and respect the relation $v_\alpha{}^i \bar{v}_j^\alpha \sim \delta_j^i$. One can expect these brackets to take form
\be\label{hypvvbr}
\big\{ v^i_\alpha, v^j_\beta  \big\} = f_1\big( v^i_\alpha v^j_\beta -  v^j_\alpha v^i_\beta  \big), \;\; \big\{ v^i_\alpha, {\bar v}_j^\beta  \big\} = f_2 \delta^i_j \delta_\alpha{}^\beta + f_3 \delta^i_j v^k_\alpha {\bar v}_k^\beta + f_4 v^i_\alpha {\bar v}_j^\beta,
\ee
where $f_1$, $f_2$, $f_3$, $f_4$ are assumed to be functions of $v^k_\gamma {\bar v}_k^\gamma$. However, it is impossible to satisfy all three demands simultaneously.

While it may seem that \p{suJeq6} is the most general solution to \p{suJeq2}, in at least one particular case exists a substantially different solution. For $n=4$, one can check that
\be\label{su4part}
J_{\alpha}{}^\beta = \im \left( z_{[\alpha\mu]}p^{[\beta\mu]} - \frac{1}{4}\delta_\alpha^\beta z_{\mu\nu}p^{\mu\nu}\right)
\ee
 satisfies equation \p{suJeq2} with nontrivial righthand side
\be\label{su4part2}
J_\alpha{}^\gamma J_\gamma{}^\beta = \frac{1}{4}\delta_\alpha{}^\beta \, J_\mu{}^\nu \, J_\nu{}^\mu \;\; \rightarrow \;\; c_1 = \frac{1}{4}, c_2 =0
\ee
and has proper brackets provided that
\be\label{zp}
\big\{ p^{\mu\nu}, z_{\alpha\beta} \big\} = \delta_{\alpha}^\mu \delta_\beta^\nu - \delta_{\alpha}^\nu \delta_\beta^\mu.
\ee
The $u(1)$ part of the $u(4)$ algebra in this case is fermionic, thus reproducing one of solutions found in \cite{kn1}. Only for this particular algebra it is possible to make $z_{\alpha\beta}$ and $p^{\mu\nu}$ real, as one can assume that
\be\label{zp2}
\big(  z_{\alpha\beta} \big)^\dagger = \frac{1}{2}\epsilon^{\alpha\beta\mu\nu}z_{\mu\nu}, \;\;  \big( p^{\mu\nu} \big)^\dagger = \frac{1}{2}\epsilon_{\mu\nu\alpha\beta}p^{\alpha\beta}.
\ee
To show that \p{su4part} satisfies equation \p{suJeq2}, one should note that due to properties of the $\epsilon^{\alpha\beta\mu\nu}$-symbol
\be\label{zp3}
z_{\alpha\gamma}p^{\beta\gamma} =-z^{\beta\gamma}p_{\alpha\gamma}+ \frac{1}{2}\delta_\alpha{}^\beta z_{\mu\nu}p^{\mu\nu}, \;\; z_{\alpha\gamma}z^{\beta\gamma} = \frac{1}{4}\delta_\alpha{}^\beta z_{\mu\nu}z^{\mu\nu}.
\ee
Let us also note that using fermionic $su(4)$ generator $\whJ_\alpha{}^\beta = \psi_\alpha\, \bpsi^\beta - \frac{1}{4}\delta_\alpha{}^\beta \psi_\mu \bpsi^\mu$ and the sum $\tJ_\alpha{}^\beta = J_\alpha{}^\beta + \whJ_\alpha{}^\beta$, one can write the appropriate Hamiltonian \p{su11N} using the Casimir operators of each algebra
\be\label{su114m}
H = \frac{1}{2}p_r^2 + \frac{1}{r^2} \left( \frac{3}{2} C_{su(4),bos} - C_{su(4)} + \frac{3}{5}C_{su(4),ferm}  \right),
\ee
where $C_{su(4),bos} = J_\mu{}^\nu J_\nu{}^\mu$, $C_{su(4),ferm} = \whJ_\mu{}^\nu \whJ_\nu{}^\mu$ and $C_{su(4)} = \tJ_\mu{}^\nu \tJ_\nu{}^\mu$.
It is worth noting that coefficients in \p{su114m} coincide with ones found in \cite{kn1} up to normalization of generators.

Therefore, the list of solutions presented here may be not exhaustive and other solutions may exist for particular algebras.

\section{Conclusion}
We constructed versions of superconformal mechanics with $SU(1,1|n)$ and $OSp(6|2)$ superconformal symmetries that involve interactions with bosonic non-Abelian currents. It was shown that for $N>4$ supersymmetry these currents have to satisfy algebraic equations. As in all the considered cases the currents are Hermitean, solutions to these equations can be found by representing the current as a diagonal matrix multiplied by a unitary matrix and its inverse. In all the considered cases unitary matrices factorize, and equations effectively constrain eigenvalues of the diagonal matrix. In the $OSp(6|2)$ case, the current is a $su(4)$ matrix with eigenvalues $(\lambda,\lambda,\lambda,-3\lambda)$ and can be expressed in terms of one row of a unitary matrix and one column  of its conjugate, which after assigning proper Dirac brackets can be treated as harmonics. The situation in the $SU(1,1|n)$ case is more complicated. In general, $u(n)$ current has two unequal eigenvalues. If one of them has multiplicity $1$, the solution can be expressed in terms of harmonics similarly to $OSp(6|2)$ case. If lowest multiplicity among the eigenvalues is two or greater, algebraic equation allows for a solution with two or greater sets of harmonics, but they are themselves algebraically constrained and no proper Dirac brackets can be assigned to them.

It is notable that obtained list of solutions is not exhaustive. For example, in the $SU(1,1|4)$ case an entirely different solution exists that does not involve harmonics but usual coordinates and momenta. This was obtained in a different representation in \cite{kn1}, but can not be explicitly found by the methods used in this article, suggesting that for particular algebras other solutions can be found. This suggestion is further strengthened by the existence of solutions of $10$ and $11$ dimensional supergravities that preserve some supersymmetry and are products of $AdS_2$ and some compact manifold. For any such solution one can expect existence of the superconformal mechanics, despite the fact that solutions found in this article (with exception of \p{su4part}) do not support such a geometric interpretation. Another related but unsolved question is the construction of the general solution for currents in $OSp(n|2)$ systems.

\end{document}